\documentclass[%
 aip,cha,
 amsmath,amssymb,
 reprint,%
]{revtex4-1}
\pdfoutput=1
\usepackage{graphicx}
\usepackage{dcolumn}
\usepackage[tight]{subfigure}
\usepackage{amsmath}
\usepackage{verbatim}
\usepackage{color}
\usepackage{bm} 
\usepackage{bbm}
\usepackage{mathbbol}
\usepackage{natbib}
\usepackage{xspace}
\usepackage{marginnote}
\usepackage{mathtools}
\usepackage{dsfont}
\usepackage{hyperref} \hypersetup{colorlinks=true,linktoc=all,linkcolor=blue,breaklinks=true,citecolor=blue,urlcolor=blue}
\usepackage{etoolbox}
\usepackage[normalem]{ulem}
\usepackage[bottom]{footmisc}
\usepackage[utf8]{inputenc}
\usepackage[T1]{fontenc}
\usepackage{mathptmx}

\robustify{\uparrow}
\robustify{\downarrow}
\robustify{\sum}
\robustify{\int}
\robustify{\nonumber}
\robustify{\cite}
\robustify{\footnote}
\newrobustcmd{\Figure}[2]{
  \begin{figure}[ht]
    \includegraphics[width=1.0\linewidth]{#1}
    \caption{#2}
  \end{figure}
}
\expandafter\newrobustcmd\csname Figure2\endcsname[3]{
\begin{figure}[ht]
  \includegraphics[width=0.9\linewidth]{#1}
  \\
  \includegraphics[width=0.9\linewidth]{#2}
  \caption{#3}
\end{figure}
}
\renewrobustcmd{\Re}{{\text{Re}}}
\renewrobustcmd{\Im}{{\text{Im}}}
\newrobustcmd{\eff}{\text{eff}} 
\newrobustcmd{\dagtot}{{\dag_\tot}}
\newrobustcmd{\dagres}{{\dag_\res}}
\newrobustcmd{\Ttot}{{\text{T}_\tot}}
\newrobustcmd{\Tres}{{\text{T}_\res}}
\newrobustcmd{\T}{{\text{T}}}
\newrobustcmd{\tot}{\text{tot}}
\newrobustcmd{\tun}{\text{T}}
\newrobustcmd{\res}{\text{R}}
\newrobustcmd{\lead}{\text{res}}
\newrobustcmd{\un}{\text{i}}   
\newrobustcmd{\In}{\text{0}}   
\newrobustcmd{\state}{inverted stationary state\xspace}   
\newrobustcmd{\K}{\mathcal{K}}
\newrobustcmd{\D}{\mathcal{I}}
\renewrobustcmd{\P}{\mathcal{P}}   
\newrobustcmd{\W}{\mathcal{W}}
\newrobustcmd{\Temp}{T}
\newrobustcmd{\dual}[1]{\bar{#1}}
\newrobustcmd{\one}{\mathds{1}}
\newrobustcmd{\ket}[1]{|#1\rangle}
\newrobustcmd{\bra}[1]{\langle#1|}
\newrobustcmd{\brkt}[1]{\langle #1 \rangle}
\newrobustcmd{\braket}[2]{\langle #1 | #2 \rangle}
\newrobustcmd{\Ket}[1]{\bm{|}#1\bm{)}}
\newrobustcmd{\Bra}[1]{\bm{(}#1\bm{|}}
\newrobustcmd{\Braket}[2]{\bm{(}#1\bm{|}#2\bm{)}}
\newrobustcmd{\Brkt}[1]{\bm{(} #1 \bm{)}}
\newrobustcmd{\op}[1]{\hat{#1}}
\newrobustcmd{\psiIn}{\psi_{\mathrm{I}}}
\newrobustcmd{\kvecIn}{\boldsymbol{k}_{\mathrm{I}}}
\newrobustcmd{\kvecInParr}{\boldsymbol{k}^{\|}_{\mathrm{I}}}
\DeclareMathOperator{\Tr}{Tr}
\newrobustcmd{\tr}{\underset{\res}{\Tr}}
\newrobustcmd{\Trlead}{\underset{\lead}{\Tr}}
\newrobustcmd{\Tralp}{\underset{\alpha}{\Tr}}
\newrobustcmd{\Trbet}{\underset{\beta}{\Tr}}
\newrobustcmd{\tri}{\Tr_\res}
\newrobustcmd{\col}[4]{{\begin{bmatrix}#1 \\ #2 \\ #3 \\ #4 \end{bmatrix}}}
\newrobustcmd{\row}[4]{{\begin{bmatrix}#1 &  #2  & #3  & #4 \end{bmatrix}}}
\newrobustcmd{\suppmat}{\cite{Schulenborg15Suppmat}}
\newrobustcmd{\Eq}[1]{Eq.~(\ref{#1})}
\newrobustcmd{\Eqs}[1]{Eqs.~(\ref{#1})}
\newrobustcmd{\BrackEq}[1]{[Eq.~(\ref{#1})]}
\newrobustcmd{\BrackApp}[1]{[App.~(\ref{#1})]}
\newrobustcmd{\BrackSec}[1]{[Sec.~(\ref{#1})]}
\newrobustcmd{\eq}[1]{(\ref{#1})}
\newrobustcmd{\Fig}[1]{Fig.~\ref{#1}}
\newrobustcmd{\BrackFig}[1]{[Fig.~\ref{#1}]}
\newrobustcmd{\fig}[1]{\ref{#1}}
\newrobustcmd{\Figs}[1]{Figs.~\ref{#1}}
\newrobustcmd{\Sec}[1]{Sec.~\ref{#1}}
\newrobustcmd{\Ref}[1]{Ref.~\citenum{#1}}
\newrobustcmd{\Refs}[1]{Refs.~\onlinecite{#1}}
\newrobustcmd{\App}[1]{App.~\ref{#1}}
\newrobustcmd{\fpop}{(-\one)^N}
\newrobustcmd{\fpOp}{(-\one)^N}
\newrobustcmd{\fpOpHat}{(-\one)^{\hat{N}}}
\newrobustcmd{\hatfpop}{(-\one)^{\hat{N}}}
\newrobustcmd{\hatfpOp}{(-\one)^{\hat{N}}}
\newrobustcmd{\zin}{z_{\text{i}}}
\newrobustcmd{\hatzin}{\hat{z}_{\text{i}}}
\newrobustcmd{\Phif}{\Phi_\mathrm{f}}
\newrobustcmd{\Phil}{\Phi_\mathrm{l}}
\newrobustcmd{\kvec}{\boldsymbol{k}}
\newrobustcmd{\kvecperp}{\boldsymbol{k}^{\perp}}
\newrobustcmd{\kvecparr}{\boldsymbol{k}^{\|}}
\newrobustcmd{\xvec}{\boldsymbol{x}}
\newrobustcmd{\pvec}{\boldsymbol{p}}
\newrobustcmd{\Vvec}{\boldsymbol{V}}
\newrobustcmd{\qZvec}{\boldsymbol{q}_0}
\newrobustcmd{\qZvecParr}{\boldsymbol{q}^{\|}_0}
\newrobustcmd{\kZvec}{\boldsymbol{k}_0}
\newrobustcmd{\uvec}{\boldsymbol{u}}
\newrobustcmd{\vvec}{\boldsymbol{v}}
\newrobustcmd{\kZvecperp}{\boldsymbol{k}^{\perp}_0}
\newrobustcmd{\kZvecalpbet}{\kvec_{\alpha\beta}}
\newrobustcmd{\kZvecalpbetParr}{\kvec^{\|}_{\alpha\beta}}
\newrobustcmd{\kZvecalpbetperp}{\kvec^{\perp}_{\alpha\beta}}
\newrobustcmd{\hOp}{\hat{h}}
\newrobustcmd{\hSC}{\hat{h}_{S}}
\newrobustcmd{\HSC}{H_{S}}
\newrobustcmd{\hW}{h_{\mathrm{W}}}
\newrobustcmd{\hWalpbet}{h_{\mathrm{W},\alpha\beta}}
\newrobustcmd{\PsiBCS}{\Psi_{BCS}}
\newrobustcmd{\HBCS}{H_{BCS}}
\newrobustcmd{\Htot}{H_{\text{tot}}}
\newrobustcmd{\hatHtot}{\hat{H}_{\text{tot}}}
\newrobustcmd{\Htun}{H_{\text{tun}}}
\newrobustcmd{\hatHtun}{\hat{H}_{\text{tun}}}
\newrobustcmd{\Hpair}{H_{\mathrm{pair}}}
\newrobustcmd{\Jens}[1]{ {\color{red}(#1)\color{black}} }
\newrobustcmd{\EBCS}{E_{BCS}}
\newrobustcmd{\EWSM}{E_{W}}
\newrobustcmd{\Csb}{C_{\mathrm{sb}}}
\newrobustcmd{\sgnfn}[1]{\mathrm{sgn}\left(#1\right)}
\newrobustcmd{\markinred}[1]{\color{red}{#1}\color{black}}
\newrobustcmd{\rsepa}{\,,\quad}
\newrobustcmd{\lsepa}{\quad ,\,}
\newrobustcmd{\lrsepa}{\quad , \quad}
\newrobustcmd{\colvec}[1]{\begin{pmatrix}#1\end{pmatrix}}
\newrobustcmd{\nbrack}[1]{\left(#1\right)}
\newrobustcmd{\sqbrack}[1]{\left[#1\right]}
\newrobustcmd{\cbrack}[1]{\left\{#1\right\}}
\newrobustcmd{\nlbrack}[1]{\left(#1\right.}
\newrobustcmd{\sqlbrack}[1]{\left[#1\right.}
\newrobustcmd{\clbrack}[1]{\left\{#1\right.}
\newrobustcmd{\nrbrack}[1]{\left.#1\right)}
\newrobustcmd{\sqrbrack}[1]{\left.#1\right]}
\newrobustcmd{\crbrack}[1]{\left.#1\right\}}
\newrobustcmd{\mean}[1]{\left\langle#1\right\rangle}
\newrobustcmd{\MeanO}[1]{\mean{#1}_0}
\newrobustcmd{\Mset}[1]{\left\{#1\right\}}
\newrobustcmd{\Mserset}[2]{\left\{#1,\dotsc,#2\right\}}
\newrobustcmd{\Forall}{\quad\forall}
\newrobustcmd{\Exists}{\quad\exists}
\newrobustcmd{\ExistsOne}{\quad\exists_1}
\newrobustcmd{\existsOne}{\exists_1}
\newrobustcmd{\MTupel}[1]{\left(#1\right)}
\newrobustcmd{\MserTup}[2]{\left(#1,\cdots,#2\right)}
\newrobustcmd{\Thetafn}[1]{\varTheta(#1)}
\newrobustcmd{\Deltafn}[1]{\delta(#1)}
\newrobustcmd{\expfn}[1]{\exp\left(#1\right)}
\newrobustcmd{\lnfn}[1]{\ln\left(#1\right)}
\newrobustcmd{\lnP}[1]{\ln_P\left(#1\right)}
\newrobustcmd{\Fermfn}[2]{\Over{\expfn{\beta\left(#1-#2\right)}+1}}
\newrobustcmd{\fermfn}[2]{f_{#2}(#1)}
\newrobustcmd{\bosefn}[2]{b_{#2}(#1)}
\newrobustcmd{\abs}[1]{\left|#1\right|}
\renewrobustcmd{\Re}{\mathrm{Re}}
\renewrobustcmd{\Im}{\mathrm{Im}}
\newrobustcmd{\rhofn}[2]{\rho_{#1}(#2)}
\newrobustcmd{\cosfn}[1]{\cos{\left(#1\right)}}
\newrobustcmd{\arcsinfn}[1]{\mathrm{arcsin}\left(#1\right)}
\newrobustcmd{\sinfn}[1]{\sin{\left(#1\right)}}
\newrobustcmd{\tanfn}[1]{\tan{\left(#1\right)}}
\newrobustcmd{\cotfn}[1]{\cot{\left(#1\right)}}
\newrobustcmd{\cothfn}[1]{\coth{\left(#1\right)}}
\newrobustcmd{\tanhfn}[1]{\tanh\!\left(#1\right)}
\newrobustcmd{\sumsub}[2]{\sum_{\substack{{#1}\\{#2}}}}
\newrobustcmd{\sumsubsub}[3]{\sum_{\substack{{#1}\\{#2}\\{#3}}}}
\newrobustcmd{\prodsub}[2]{\prod_{\substack{{#1}\\{#2}}}}
\newrobustcmd{\cc}[1]{\overline{#1}}
\newrobustcmd{\equalby}[1]{\overset{#1}{=}}
\newrobustcmd{\equalbys}[2]{\underset{#2}{\overset{#1}{=}}}
\newrobustcmd{\equalbyeqn}[1]{\overset{\eqref{#1}}{=}}
\newrobustcmd{\equalbyeqns}[2]{\underset{\eqref{#2}}{\overset{\eqref{#1}}{=}}}
\newrobustcmd{\evalAtEqui}[1]{#1\big|_{\text{eq}}}
\newrobustcmd{\evalAtBal}[1]{\left.#1\right|_{\text{bal}}}
\newrobustcmd{\zeq}{z_{\text{eq}}}
\newrobustcmd{\hatzeq}{\hat{z}_{\text{eq}}}
\newrobustcmd{\zieq}{z_{\text{i,eq}}}
\newrobustcmd{\hatzieq}{\hat{z}_{\text{i,eq}}}
\newrobustcmd{\zia}{z_{\text{i}\alpha}}
\newrobustcmd{\za}{z_{\alpha}}
\newrobustcmd{\hatzia}{\hat{z}_{\text{i}\alpha}}
\newrobustcmd{\nzia}{n_{\text{i}\alpha}}
\newrobustcmd{\nziap}{n_{\text{i}\alpha'}}
\newrobustcmd{\pzia}{p_{\text{i}\alpha}}
\newrobustcmd{\Eeq}{E_{\text{eq}}}
\newrobustcmd{\nza}{n_{z\alpha}}
\newrobustcmd{\pza}{p_{z\alpha}}
\newrobustcmd{\nzap}{n_{z\alpha'}}
\newrobustcmd{\nz}{n_{z}}
\newrobustcmd{\nzeq}{n_{z,\text{eq}}}
\newrobustcmd{\delnsqzeq}{\delta n^2_{z,\text{eq}}}
\newrobustcmd{\pzeq}{p_{z,\text{eq}}}
\newrobustcmd{\pzieq}{p_{i,\text{eq}}}
\newrobustcmd{\ceq}{c_{\text{eq}}}
\newrobustcmd{\cpeq}{c'_{\text{eq}}}
\newrobustcmd{\peq}{p_{\text{eq}}}
\newrobustcmd{\ppeq}{p'_{\text{eq}}}
\newrobustcmd{\gamc}{\gamma_c}
\newrobustcmd{\gamp}{\gamma_p}
\newrobustcmd{\gamca}{\gamma_{c\alpha}}
\newrobustcmd{\gampa}{\gamma_{p\alpha}}
\newrobustcmd{\gamceq}{\gamma_{c,\text{eq}}}
\newrobustcmd{\gamceff}{\gamma^{\text{eff}}_{c}}
\newrobustcmd{\gamceffa}{\gamma^{\text{eff}}_{c\alpha}}
\newrobustcmd{\gamceffeq}{\gamma^{\text{eff}}_{c,\text{eq}}}
\newrobustcmd{\gampeff}{\gamma^{\text{eff}}_{p}}
\newrobustcmd{\gampeffa}{\gamma^{\text{eff}}_{p\alpha}}
\newrobustcmd{\gampeffeq}{\gamma^{\text{eff}}_{p,\text{eq}}}
\newrobustcmd{\gamceps}{\gamma_{c\epsilon}}
\newrobustcmd{\gamcepsa}{\gamma_{c\epsilon,\alpha}}
\newrobustcmd{\gamcepseq}{\gamma_{c\epsilon,\text{eq}}}
\newrobustcmd{\gamcU}{\gamma_{cU}}
\newrobustcmd{\gamcUa}{\gamma_{cU,\alpha}}
\newrobustcmd{\gamcUeq}{\gamma_{cU,\text{eq}}}
\newrobustcmd{\gameq}{\gamma_{\text{eq}}}
\newrobustcmd{\gampeq}{\gamma_{p,\text{eq}}}
\newrobustcmd{\gamci}{\dual{\gamma}_c}
\newrobustcmd{\gampi}{\dual{\gamma}_p}
\newrobustcmd{\gamcia}{\dual{\gamma}_{c\alpha}}
\newrobustcmd{\gampia}{\dual{\gamma}_{p\alpha}}
\newrobustcmd{\gamcieq}{\dual{\gamma}_{c,\text{eq}}}
\newrobustcmd{\gamceffi}{\dual{\gamma}^{\text{eff}}_{c}}
\newrobustcmd{\gamceffia}{\dual{\gamma}^{\text{eff}}_{c\alpha}}
\newrobustcmd{\gamceffieq}{\dual{\gamma}^{\text{eff}}_{c,\text{eq}}}
\newrobustcmd{\gampeffi}{\dual{\gamma}^{\text{eff}}_{p}}
\newrobustcmd{\gampeffia}{\dual{\gamma}^{\text{eff}}_{p\alpha}}
\newrobustcmd{\gampeffieq}{\dual{\gamma}^{\text{eff}}_{p,\text{eq}}}
\newrobustcmd{\gamcepsi}{\dual{\gamma}_{c\epsilon}}
\newrobustcmd{\gamcepsia}{\dual{\gamma}_{c\epsilon,\alpha}}
\newrobustcmd{\gamcepsieq}{\dual{\gamma}_{c\epsilon,\text{eq}}}
\newrobustcmd{\gamcUi}{\dual{\gamma}_{cU}}
\newrobustcmd{\gamcUia}{\dual{\gamma}_{cU,\alpha}}
\newrobustcmd{\gamcUieq}{\dual{\gamma}_{cU,\text{eq}}}
\newrobustcmd{\gamieq}{\dual{\gamma}_{\text{eq}}}
\newrobustcmd{\gampieq}{\dual{\gamma}_{p,\text{eq}}}
\newrobustcmd{\nzieq}{n_{\text{i,eq}}}
\newrobustcmd{\delnsqzieq}{\delta n^2_{\text{i,eq}}}
\newrobustcmd{\delnsqzia}{\delta n^2_{\text{i}\alpha}}
\newrobustcmd{\delnsqzi}{\delta n^2_{\text{d}}}
\newrobustcmd{\delnsqz}{\delta n^2_{z}}
\newrobustcmd{\delnsqza}{\delta n^2_{z\alpha}}
\newrobustcmd{\nzi}{n_{\text{d}}}
\newrobustcmd{\pzi}{p_{\text{i}}}
\newrobustcmd{\kBT}{k_{\text{B}}T}
\newrobustcmd{\kB}{k_{\text{B}}}
\newrobustcmd{\Hdot}{H}
\newrobustcmd{\Hleads}{H_{\text{res}}}
\newrobustcmd{\Hlead}{H_{\text{res}}}
\newrobustcmd{\hatHleads}{\hat{H}_{\text{R}}}
\newrobustcmd{\INa}{I_{\text{N}}^{\alpha}}
\newrobustcmd{\INap}{I_{\text{N}}^{\alpha'}}
\newrobustcmd{\IEa}{I_{\text{E}}^{\alpha}}
\newrobustcmd{\GamR}{\Gamma_{\text{R}}}
\newrobustcmd{\GamL}{\Gamma_{\text{L}}}
\newrobustcmd{\muR}{\mu_{\text{R}}}
\newrobustcmd{\muL}{\mu_{\text{L}}}
\newrobustcmd{\mua}{\mu_{\alpha}}
\newrobustcmd{\TR}{T_{\text{R}}}
\newrobustcmd{\TL}{T_{\text{L}}}
\newrobustcmd{\Ta}{T_{\alpha}}
\newrobustcmd{\IR}{I^{\text{R}}}
\newrobustcmd{\IL}{I^{\text{L}}}
\newrobustcmd{\INR}{I^{\text{R}}_N}
\newrobustcmd{\INL}{I^{\text{L}}_N}
\newrobustcmd{\IER}{I^{\text{R}}_E}
\newrobustcmd{\IEL}{I^{\text{L}}_E}
\newrobustcmd{\JR}{J^{\text{R}}}
\newrobustcmd{\JL}{J^{\text{L}}}
\newrobustcmd{\mueq}{\mu}
\newrobustcmd{\sudag}{{\boldsymbol{\dagger}}}
\newrobustcmd{\N}{\mathds{N}}
\newrobustcmd{\Gamop}{\mathbb{\Gamma}}
\newrobustcmd{\Gamopa}{\mathbb{\Gamma}_{\alpha}}
\newrobustcmd{\cP}{\mathcal{P}}
\newrobustcmd{\Gamepsa}{\Gamma_{\epsilon\alpha}}
\newrobustcmd{\GamepsL}{\Gamma_{\epsilon\text{L}}}
\newrobustcmd{\GamepsR}{\Gamma_{\epsilon\text{R}}}
\newrobustcmd{\Gamepsap}{\Gamma_{\epsilon\alpha'}}
\newrobustcmd{\GamUa}{\Gamma_{U\alpha}}
\newrobustcmd{\Gama}{\Gamma_{\alpha}}
\newrobustcmd{\GamUL}{\Gamma_{U\text{L}}}
\newrobustcmd{\GamUR}{\Gamma_{U\text{R}}}
\newrobustcmd{\GamUap}{\Gamma_{U\alpha'}}
\newrobustcmd{\Gameps}{\Gamma_{\epsilon}}
\newrobustcmd{\Repsaap}{R^{\alpha\alpha'}_{\epsilon}}
\newrobustcmd{\RepsLL}{R^{\text{LL}}_{\epsilon}}
\newrobustcmd{\RepsLR}{R^{\text{LR}}_{\epsilon}}
\newrobustcmd{\RepsRL}{R^{\text{RL}}_{\epsilon}}
\newrobustcmd{\RepsRR}{R^{\text{RR}}_{\epsilon}}
\newrobustcmd{\RUaap}{R^{\alpha\alpha'}_{U}}
\newrobustcmd{\RULL}{R^{\text{LL}}_{U}}
\newrobustcmd{\RULR}{R^{\text{LR}}_{U}}
\newrobustcmd{\RURL}{R^{\text{RL}}_{U}}
\newrobustcmd{\RURR}{R^{\text{RR}}_{U}}
\newrobustcmd{\GamU}{\Gamma_{U}}
\newrobustcmd{\nph}{n_{\text{ph}}}
\newrobustcmd{\nphz}{n_{\text{ph},z}}
\newrobustcmd{\Snl}{S_{\text{nl}}}
\newrobustcmd{\delEps}{\epsilon}
\newrobustcmd{\Pmax}{P_{\text{max}}}
\newrobustcmd{\etaPmax}{\eta_{P_{\text{max}}}}
\makeatletter
\AtBeginDocument{\let\LS@rot\@undefined}
\makeatother

\begin{document}

\preprint{AIP/123-QED}

\title[Thermovoltage in quantum dots with attractive interaction]{Thermovoltage in quantum dots with attractive interaction}

\author{Jens Schulenborg}
\email[Corresponding author: ]{jens.schulenborg@nbi.ku.dk}
\affiliation{Center for Quantum Devices, Niels Bohr Institute, University of Copenhagen, 2100 Copenhagen, Denmark}
\affiliation{Department of Microtechnology and Nanoscience, Chalmers University of Technology, 41296 G{\"o}teborg, Sweden}
\author{Maarten R. Wegewijs}%
\affiliation{ Institute for Theory of Statistical Physics,
      RWTH Aachen,52056 Aachen, Germany}
\affiliation{
 Peter Gr{\"u}nberg Institut and JARA,
      Forschungszentrum J{\"u}lich, 52425 J{\"u}lich, Germany}%
\author{Janine Splettstoesser}
\affiliation{Department of Microtechnology and Nanoscience, Chalmers University of Technology, 41296 G{\"o}teborg, Sweden}

\date{\today}

\begin{abstract}
We study the linear and nonlinear thermovoltage of a quantum dot with effective \textit{attractive} electron-electron interaction and weak, energy-dependent tunnel-coupling to electronic contacts. Remarkably, we find that the thermovoltage shows signatures of \textit{repulsive} interaction which can be rationalized. These thermovoltage characteristics are robust against large potential and temperature differences well into the nonlinear regime, which we expect can be demonstrated in current state-of-the-art experiments. Furthermore, under nonlinear operation, we find extended regions of large power production  at efficiencies on the order of the  Curzon-Ahlborn bound interrupted only by a characteristic sharp dip. 
\end{abstract}

\maketitle

Recently, different types of devices with an effectively attractive electron-electron interaction~\cite{Butler2015Mar} have been experimentally investigated~\cite{Cheng2015May} and also quantum dot structures with attractive onsite-interaction have been realized~\cite{Hamo2016Jul,Prawiroatmodjo2017Aug}. In these quantum dots, signatures of pair-tunneling~\cite{Koch2006Feb,Sela2008Feb,Leijnse2009Oct} induced by the attractive onsite interaction  could be identified in transport properties.

In the present letter, we predict surprising features in the thermovoltage  of such quantum dots.
We show that the linear-response thermovoltage --- the Seebeck coefficient --- shows signatures at quantum-dot level positions that are characteristic for the Coulomb-oscillations due to \textit{repulsive} onsite interaction.
We rationalize this fact and show it can be exploited in an analysis which demonstrates
how these features are modified under various, realistic experimental conditions. 
The discussed effects are highly relevant for the characterization of attractive systems, which has only recently started. 

Simultaneously, there has been significant progress in investigating linear and nonlinear-response thermoelectrics in quantum dot devices, see \Refs{Esposito2009Apr,Leijnse2010Jul,Sanchez2011Feb,Svensson2013Oct,Kennes2013Jun,Sothmann2014Dec,Svilans2016Aug,Sanchez2016Dec,Schulenborg2017Dec,Walldorf2017Sep,Erdman2017Jun,Schulenborg2018Dec,Josefsson2018Jul} and references therein. 
These are of interest for on-chip energy harvesting and their Seebeck coefficient is a key parameter to characterize them. Our analysis instead shows the thermoelectric properties of systems  with \emph{strong attractive} electron-electron interaction. We also explicitly address energy-dependent tunnel couplings between dot and environment as energy filters in addition to the quantum-dot levels. Efficient nanoscale thermoelectrics, in particular three-terminal energy harvesters~\cite{Sanchez2011Feb,Hartmann2015Apr,Roche2015Apr,Thierschmann2015Aug,Thierschmann2016Dec}, crucially rely on this energy-dependent coupling.
In this letter, we characterize the performance of such quantum dots with attractive interaction as steady-state heat engines and find extended regions of large power production and efficiency.

Finally, the thermoelectric response of \emph{repulsive} quantum dots~\cite{Schulenborg2017Dec,Schulenborg2018Dec} has successfully been analyzed with a mapping based on a fermionic duality relation~\cite{Schulenborg2016Feb}, providing simple analytic formulas.
Here, this relation enables us to explain the thermoelectric response of a quantum dot with attractive interaction in terms of the well understood physics of a repulsive dot. This simple description can serve as a guide for future experiments.

\begin{figure}
\includegraphics[width=\linewidth]{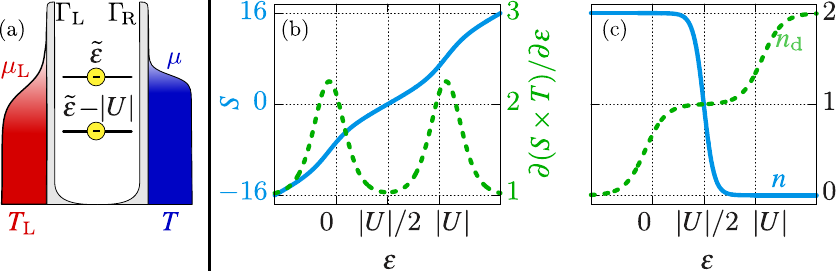}
\caption{\label{fig_linearWB}
(a) Sketch of a quantum dot with attractive interaction. (b) Seebeck coefficient $S$ (blue solid) and its derivative (green dashed), and (c) equilibrium charge $n$ and its dual $\nzi$, as function of dot level $\epsilon = \tilde{\epsilon} - \mu$. We take $T=|U|/10$ and $\Gamma_\text{L,R}$ energy-independent. 
}
\end{figure}

The quantum dot of interest is sketched in \Fig{fig_linearWB}(a). It is modelled as a single spin-degenerate level, with an \textit{attractive} electron-electron interaction. We assume the level spacing to be large compared to any other energy scale relevant for transport, such as voltage bias and temperatures; indeed, recent experimental realizations of quantum dots with attractive interactions have been well explained in terms of such a model~\cite{Prawiroatmodjo2017Aug}. The isolated dot is then described by the Hamiltonian
\begin{equation}
H = \tilde{\epsilon} N - |U| N_\uparrow N_\downarrow\label{eq_hamiltonian}
\end{equation}
with the single-level energy $\tilde{\epsilon}$, the local particle-number operator $N = N_\uparrow + N_\downarrow$ with spin-resolved components $N_\uparrow$ and $N_\downarrow$, and the interaction strength $U=-|U|$. 
The quantum dot is tunnel-coupled
to two electronic reservoirs, $\alpha = \text{L,R}$, with electrochemical potentials $\mu_\text{L}=\mu-V$ and $\mu_\text{R}=\mu$, and temperatures $T_\text{L}=T+\Delta T$ and $T_\text{R}=T$.
We assume the experimentally relevant case of  weak  tunnel rates $\Gamma_\alpha(E) \ll T_{\alpha}$, implying that pair-tunneling is enabled by thermal excitations~\cite{Prawiroatmodjo2017Aug,Kleinherbers2018Jul,Placke2018Aug}. We allow these rates to strongly depend on the energy of the tunneling process.
We set $k_\text{B} = \hbar = |e| = 1$.

We analyze the full thermoelectric response of the quantum dot using expressions obtained from a recently established, general \emph{fermionic duality}~\cite{Schulenborg2016Feb,Schulenborg2017Dec,Schulenborg2018Dec,Bruch2020}.
This purely dissipative symmetry, applied to a  weak-coupling master-equation description,
maps the transport dynamics of fermionic open non-equilibrum systems to those of \emph{dual} systems with sign-inverted local energies, chemical potentials, and energy-dependencies of the tunnel couplings,
replacing
$[\tilde{\epsilon},-|U|,\mu_\alpha,\Gamma_\alpha(E)] \rightarrow [-\tilde{\epsilon},|U|,-\mu_\alpha,\Gamma_\alpha(-E)]$.
In the present case, the remarkable duality enables us to predict \emph{non-equilibrium} effects in the thermoelectric response of the attractive dot of interest by relating them to the properties of  the quantum dot with \textit{equilibrium} parameters and attractive interaction as well as to its dual model with \emph{repulsive} interaction~\footnote{Up to linear order in the couplings $\Gamma_\alpha$, the dual model is a physically valid model and can be interpreted physically. This breaks down in higher-than-linear orders in $\Gamma_\alpha$, but the duality remains useful~\cite{Bruch2020}.}. The key quantities in this analysis are the equilibrium dot occupation for attractive interaction
\begin{equation}
 n(\epsilon,-|U|)  =   \langle N\rangle = \frac{2f(\epsilon)}{1 + f(\epsilon) - f(\epsilon - |U|)},\label{eq_charge}
 \end{equation}
 with $\epsilon = \tilde{\epsilon} - \mu$ and  Fermi function $f(x) = \sqbrack{\expfn{\frac{x}{T}} + 1}^{-1}$, and most importantly, the \emph{dual} occupation~\cite{Schulenborg2018Dec},
 \begin{equation}
 \nzi  =  n(-\epsilon,|U|)= \frac{2\left[1-f(\epsilon)\right]}{1 - f(\epsilon) + f(\epsilon - |U|)},\label{eq_dual_charge}
\end{equation}
to which the duality assigns a repulsive interaction $|U|$. Figure~\ref{fig_linearWB}(c) shows both $n$ and $\nzi$ as functions of $\epsilon$.

We start with the linear response of the thermovoltage, for small $V$ and $\Delta T$, the linear Seebeck coefficient $S=V/\Delta T|_{I = 0}$ at vanishing charge current $I = 0$ (see \Eq{eq_current}) across the dot, and  first consider energy-independent tunnel couplings $\Gamma_{\alpha}(E) \rightarrow \Gamma_{\alpha}$. 
The explicit formula for $S$ has a remarkably simple form~\cite{Schulenborg2017Dec} in terms of the dual dot occupation \eq{eq_dual_charge},
\begin{equation}
S\times T = \delEps - |U|(2 - \nzi)/2.\label{eq_seebeck_wideband}
\end{equation}
The consequences of the attractive interaction are shown in Figure~\ref{fig_linearWB}(b). On the one hand, we find a linear $\epsilon$-dependence $S\times T \approx \delEps -|U|/2$ around the zero-crossing at $\delEps = |U|/2$. This is intuitively expected in analogy to the well-known case of repulsive quantum dots~\cite{Beenakker1992Oct,Staring1993Apr,Dzurak1993Sep,Schulenborg2017Dec}: as reflected by $n(\epsilon)$ in \Fig{fig_linearWB}(c), the attractive dot effectively acts as a single resonance at $\delEps = |U|/2$. 
On the other hand, we find $S \times T\rightarrow \delEps - |U|$ for $\delEps < 0$ and $S \times T\rightarrow \delEps$ for $\delEps > |U|$. 
Here, the attractive interaction does not anymore favor thermally excited pair transitions over single-electron transitions. 
The resulting crossovers between all three identified regimes lead to surprising kinks in the Seebeck coefficient $S(\epsilon)$ [blue line in \Fig{fig_linearWB}(b)]. These kinks are even better visible in its derivative~\footnote{The relation $\delnsqzi = T\partial\nzi/\partial\epsilon$ is proven in \Ref{Schulenborg2017Dec}.} [green, dashed line in  \Fig{fig_linearWB}(b)], measurable using lock-in techniques
\begin{equation}
 T\times \frac{\partial S}{\partial\epsilon} = 1 + \delnsqzi\times\frac{|U|}{2T}.
 \label{eq_Sder}
 \end{equation}
Indeed, this derivative depends on the equilibrium charge fluctuations $\delnsqzi = \langle N^2\rangle_{\text{d}} - \nzi^2$ after the duality mapping. 
This implies features in $S(\epsilon)$ at the Coulomb resonances $\delEps = 0$ and $\delEps = -U =|U|$ of a \emph{repulsive} dot. 
Obtaining the dual occupation $\nzi$ in \Eq{eq_seebeck_wideband} by a brute-force ``Fermi's Golden rule'' calculation is surprising and defies common physical intuition. This is typical for insights offered by fermionic duality~\cite{Schulenborg2016Feb,Schulenborg2017Dec,Schulenborg2018Dec}.

The mapping to a repulsive system via $\nzi$ enables a further prediction for experiments. The peaks in $\partial S/\partial\epsilon$ are shifted by approximately $\pm T\ln(2)$ away from the zero-temperature resonances $\delEps = 0,|U|$ of a repulsive dot. In the latter case, this shift is well understood as the entropy of the singly-occupied state due to the spin degeneracy~\cite{Deshmukh2002Jan,Bonet2002Jan,Juergens2013Jun}. In an attractive dot, single occupation is never a stable equilibrium state, but remarkably, our dual picture reveals that its spin degeneracy nevertheless affects the Seebeck coefficient.

\begin{figure}
\includegraphics[width=1.0\linewidth]{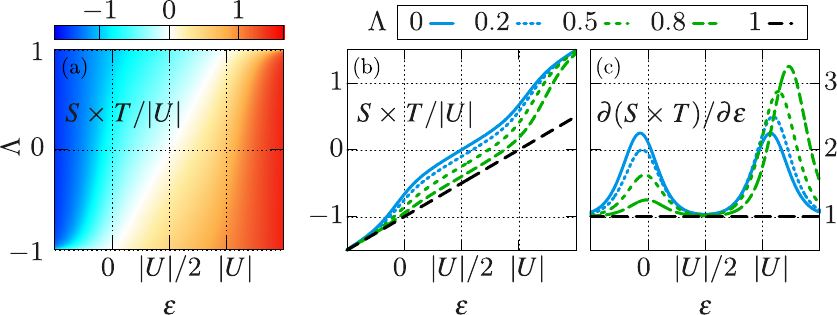}
\caption{\label{fig_linearNonWB}
Seebeck coefficient [(a) and (b)] and its derivative at fixed $\Lambda$ (c) as function of dot level $\epsilon$ and coupling-asymmetry $\Lambda$. We take energy-dependent $\GamL(E),\GamR(E)$ and $T=|U|/10$. 
}
\end{figure}

A relevant question is how  energy-dependent couplings $\Gamma_\alpha(E)$ affect the thermovoltage. In experiments, the environment density of states may sizably vary around the Fermi energy and thereby give rise to such an energy dependency. Moreover, an appropriately tuned energy dependence can be beneficial for efficient nanoscale energy harvesting~\cite{Sanchez2011Feb,Hartmann2015Apr,Thierschmann2015Aug,Roche2015Apr,Walldorf2017Sep}. 

We account for this by assuming arbitrary smoothly energy-dependent rates $\Gamma_{\alpha=\text{L,R}}(E)\ll T$ within the weak-coupling constraint.
Following \Ref{Schulenborg2018Dec}, $S$ is then determined by
\begin{align}
S\times T &= \delEps - \frac{(1 + \Lambda)(2 - \nzi)}{(1 - \Lambda)\nzi + (1 + \Lambda)(2 - \nzi)}|U|\label{eq_linearNonWB}. 
\end{align}
This introduces the energy asymmetry $\Lambda$ of the coupling,
\begin{equation}
\Lambda = \frac{\GamUL\GamUR\Gameps - \GamepsL\GamepsR\GamU}{\GamUL\GamUR\Gameps + \GamepsL\GamepsR\GamU},
\label{eq_coupling_asymmetry}
\end{equation}
with $\GamepsL = \GamL(\epsilon)$, $\GamepsR = \GamR(\epsilon)$, $\GamUL = \GamL(\epsilon -|U|)$, $\GamUR = \GamR(\epsilon - |U|)$, $\Gameps = \GamepsL + \GamepsR$, and $\GamU = \GamUL + \GamUR$.
The result for $S$ in the presence of energy-dependent tunnel coupling is shown in \Fig{fig_linearNonWB}. 
Equation~(\ref{eq_linearNonWB}) enables us to systematically isolate how energy-dependent couplings $\Gamma_\alpha(E)$ influence the linear thermovoltage for different level positions $\epsilon$ at fixed $U =-|U|$ and $T$~\footnote{$\Lambda$ actually depends on $\epsilon$ via $\Gamma_\alpha(E)$ \BrackEq{eq_coupling_asymmetry}. In \Fig{fig_linearNonWB}, $\epsilon$ is swept while adjusting $\Gamma_\alpha(E)$ to keep $\Lambda$ constant. However, as \Ref{Schulenborg2018Dec} has shown, an $\epsilon$-sweep that keeps $\Lambda$ constant is achieved without adjusting the couplings for an exponential $\Gamma_\alpha(E) \sim \expfn{(E - E_0)/D}$. Such an energy-profile has proven to be a reasonable assumption in analyzing several past experiments on few-electron transport from quantum dots~\cite{Matveev1996Oct,Zimmerman2004Nov,Fletcher2012Oct,Kaestner2015Sep}.}.
In \Fig{fig_linearNonWB}, we identify---as one main qualitative effect of energetic coupling asymmetry---a shift of the zero-crossing of $S$ as a function of $\epsilon$ away from $\epsilon_0  = |U|/2$. This shift, which we call $\Delta\epsilon_0 $,  can be understood from \Eq{eq_linearNonWB}. We exploit that well within $0 < \delEps < |U|$ and for $|U| \gg T$, repulsive Coulomb blockade induces a plateau at $\nzi \approx 1$ in the dual occupation. This simplifies \Eq{eq_linearNonWB} to
\begin{equation}
 S\times T \rightarrow \delEps - (1 + \Lambda)|U|/2\label{eq_seebeck_shift} \quad\text{ for }\quad 0 < \epsilon < |U|,
\end{equation}
implying that an asymmetry $\Lambda > 0$ favors emission from a doubly occupied dot at addition energy $\delEps - |U|$, and $\Lambda < 0$ favors absorption into an empty dot at energy $\delEps$. This offset of $S(\epsilon)$ in \Eq{eq_seebeck_shift} involves a zero crossing $\epsilon_0$ shifted away from $|U|/2$ by $\Delta\epsilon_0 \rightarrow |U|/2\times\Lambda$, predicting a pronounced effect for strong attractive interaction. The limit $\Delta\epsilon_0 \rightarrow \pm |U|/2$ for $\Lambda \rightarrow \pm 1$ reflects that transport is reduced to a single resonance at $\delEps = 0$ or $\delEps = |U|$, annulling all two-electron features.

Next, \Fig{fig_linearNonWB}(c) demonstrates how the peaks in $\partial S/\partial\epsilon$, and hence the kinks in $S$ visible in \Fig{fig_linearWB}(b), change with energy-dependent couplings. For $\Lambda > 0$, the peak in $\partial S/\partial\epsilon$ around $\delEps = 0$ shrinks with larger $|\Lambda|$, whereas the peak in $\partial S/\partial\epsilon$ around $\delEps =  |U|$ grows. At the same time, the latter also moves substantially further away from resonance---we refer to this shift from resonance as $\Delta\epsilon_{\text{P}}$. For $\Lambda < 0$, the behavior is opposite with respect to the two peaks. In the single-resonance limit $|\Lambda| \rightarrow 1$, the slope $\partial S/\partial\epsilon \times T \rightarrow 1$ becomes constant and both peaks disappear, as expected.

The change in relative peak height follows from the offset of $S$ within $0 < \delEps < |U|$ described in \Eq{eq_seebeck_shift}.  For, e.g., $\Lambda > 0$, the shift to smaller $S$ in this $\epsilon$-interval causes the step of $S(\epsilon)$ around $\delEps = 0$ to be smaller than around $\delEps = |U|$, see \Fig{fig_linearNonWB}(b). This leads to a smaller relative peak height in $\partial S/\partial\epsilon$. 

The growing shift $\Delta\epsilon_{\text{P}}$ of the higher peak for increasing $|\Lambda|$ stems from the fact that 
the coupling asymmetry $\Lambda$ not only affects $S$ for $0 < \delEps < |U|$ and large $|U|/T$ as in \Eq{eq_seebeck_shift}, but in general influences where the crossover between the single-particle, linear limits [$S(\epsilon) \sim \delEps$ and $S(\epsilon) \sim \delEps -|U|$] and the  two-particle limit [$S(\epsilon) \sim \delEps-|U|/2$] takes place, see \Eq{eq_linearNonWB}. 
For both small $|\Lambda|$ and large $|\Lambda| \lesssim 1$ a useful analytical expression is
(including the spin-degeneracy shift $T\ln(2)$)
\begin{equation}
\Delta\epsilon_{\text{P}} \approx T\times\sgnfn{\Lambda}\times\ln\sqbrack{2/(1 - |\Lambda|)}.\label{eq_peak_shift}
\end{equation}
For example, $\Lambda = 0.8$ yields $\Delta\epsilon_{\text{P}} \approx 3.3 T\ln(2)$, which substantially deviates from the wideband limit result, where the shift away from resonance is given by $T\ln(2)$. In this case, $\GamU$ is large enough compared to $\Gameps$, such that  even for a considerable interval with $\delEps > 0$ and $\delEps - |U| > 0$, the physics of pair tunneling prevails. Namely,  a thermally excited electron entering the dot at energy $\epsilon$ causes transport of further electrons at energy $\epsilon - |U|$ before the dot is emptied again. 

\begin{figure}
\includegraphics[width=\linewidth]{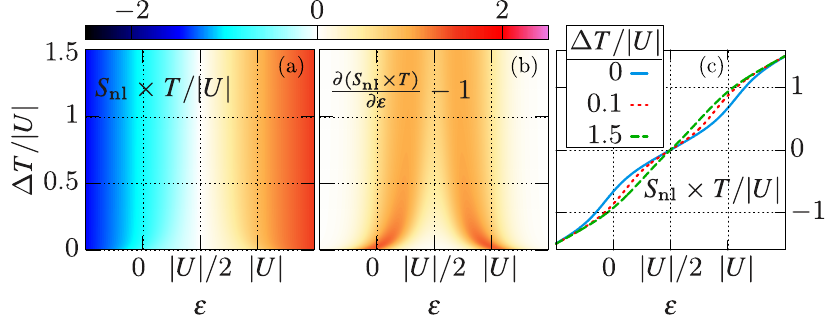}
\caption{\label{fig_nonlinear}
Nonlinear thermovoltage [(a) and (c)] and its derivative (b) as function of dot level $\epsilon$ and temperature gradient $\Delta T/|U|$. We take $T=|U|/10$ and $\GamL,\GamR$ energy-independent. 
}
\end{figure}

Next, we demonstrate how the Seebeck coefficient gets modified in the \emph{nonlinear regime} due to large $\Delta T$ and $V$. This is also relevant below where we discuss the performance of the quantum dot as a thermoelectric device.
The nonlinear thermovoltage $\Snl = V|_{I = 0}/\Delta T$ quantifies the voltage $V = \mu - \muL$ required to suppress a charge current $I$ induced by a large temperature difference $\Delta T = \TL - T$ across the junction. 

We have previously shown~\cite{Schulenborg2017Dec} the nonlinear current to assume the compact form
\begin{equation}
 I = \frac{\gamma_{\text{L}}\gamma_{\text{R}}}{\gamma_{\text{L}}+\gamma_{\text{R}}}\nbrack{n_{\text{L}} - n_{\text{R}}}\label{eq_current}
\end{equation}
in the wideband limit $(\Gamepsa = \GamUa = \Gama)$. This depends on the difference between \emph{equilibrium} occupations $n_{\text{R}} = n$ and $n_{\text{L}} = \left.n\right|_{\mu,T\rightarrow \muL,\TL}$, and on the energy-level dependent charge relaxation rates~\cite{Splettstoesser2010Apr,Vanherck2017Mar} $\gamma_{\text{R/L}} = \Gamma_{\text{R/L}}\sqbrack{1 + f_{\text{R/L}}(\epsilon) - f_{\text{R/L}}(\epsilon - |U|)}/2 > 0$. Both the occupations and relaxation rates can be understood as if the dot was coupled only to the right  or left  lead; the symbols $f_{\text{R}}(x) = f(x)$ and $f_{\text{L}}(x) = \left.f(x)\right|_{\mu,T\rightarrow\muL,\TL}$ denote the corresponding Fermi functions.  
Setting $I = 0$ while keeping the potential $\muR = \mu$ and temperature $\TR = T$ fixed, \Eq{eq_current} yields the helpful analytical result for the thermovoltage~\cite{Schulenborg2017Dec}
\begin{align}
 &\Snl\times T = \delEps - |U|\label{eq_nonlinear}\\
 &- \frac{T + \Delta T}{\Delta T/T}\ln\sqbrack{\frac{1 - \nzi + \sqrt{(1 - \nzi)^2 +e^{\frac{-|U|\Delta T}{T(T + \Delta T)}}\nzi(2-\nzi)}}{2 - \nzi}}\notag
\end{align}
again expressed in terms of the dual occupation number $\nzi$. 

Figure~\ref{fig_nonlinear} shows $\Snl \times T$ and its $\epsilon$-derivative~\footnote{The derivative can be straightforwardly analytically obtained  from \Eq{eq_nonlinear}.} as a function of level position $\epsilon$ and temperature difference $\Delta T > 0$. As expected, the zero crossing at the particle-hole symmetric point $\delEps = |U|/2$ persists. Importantly,  the counter intuitive features at $\delEps = 0,|U|$ also continue to exist. This is indeed suggested by \Eq{eq_nonlinear}, in which the $\epsilon$-dependence enters \textit{entirely} through the dual occupation $\nzi$ determined by repulsive interaction. 
Specifically, in Fig.~\ref{fig_nonlinear}(c), an increasing $\Delta T $ transforms the steps at $\delEps = 0,|U|$ between the three regimes of $\Snl(\epsilon)\times T$ with equal $\epsilon$-slopes into temperature-broadened transitions between three regimes of \emph{different} slopes. For $\delEps < 0$ and $\delEps > |U|$, $\Snl(\epsilon)\times T$ still grows with slope $1$ as function of $\delEps$, just as the Seebeck coefficient $S(\epsilon)\times T$, see Fig.~\ref{fig_linearWB}(b). This again reflects that transport is effectively governed by single-particle physics, see above discussion on energy-dependent couplings.
For levels $0 < \delEps < |U|$ at which two-particle effects are relevant, a linear $\epsilon$-dependence of $\Snl\times T$ with larger slope $\sim 2$ emerges, as can be qualitatively understood from an analysis of the nonlinear charge current \eq{eq_current}: a small $T \ll |U|$ and a large $\Delta T \gtrsim |U|$ corresponds to a sharp two-particle transition of $n_{\text{R}} = 2 \rightarrow 0$, yet a smooth behavior of $n_{\text{L}}$  as a function of $\epsilon$ around $\delEps = |U|/2$. Consequently, fulfilling $n_{\text{L}} = n_{\text{R}}$ to achieve $I = 0$ for fixed $\epsilon$ and $\mu$ requires a relatively large shift of $V = \mu - \muL$. In particular, the slope $2$ of $\Snl(\epsilon)\times T = \left.\frac{VT}{\Delta T}\right|_{I=0}$ in the limit $\Delta T/T\gg1$ reflects the sharp change of $n_{\text{R}}$ by $2$ at $\epsilon = |U|/2$ due to attractive interaction, see  \Fig{fig_linearWB}(c). 

Finally, let us consider the power output of the dot. As well-known~\cite{Mahan1996Jul,Hicks1993May,Hicks1993Jun,Humphrey2005Mar}, a sharp spectral resonance of a conductor is beneficial for its thermoelectric performance. Hence, quantum dots have been studied as thermoelectric elements operated in the nonlinear regime of large temperature and voltage biases, both theoretically~\cite{Leijnse2010Jul,Esposito2009Apr,Kennes2013Jun,Schulenborg2017Dec} and experimentally~\cite{Svensson2013Oct,Josefsson2018Jul}. We now show that also in the presence of strong attractive interaction, finite power output is possible at high efficiencies.

\begin{figure}
\includegraphics[width=\linewidth]{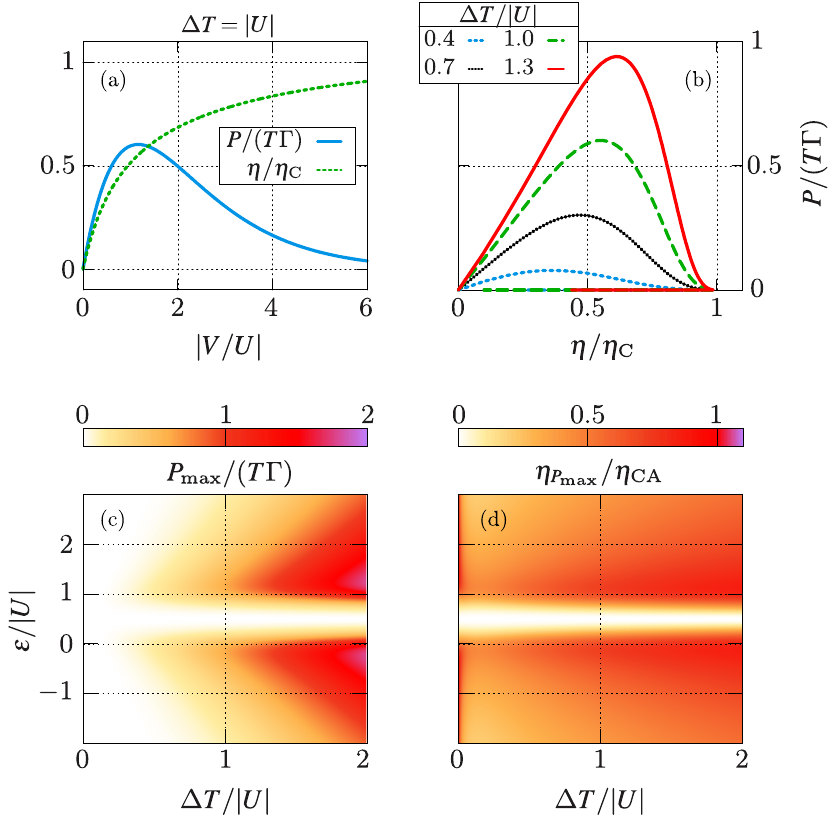}
\caption{\label{fig_performance}
(a) Power $P$ and efficiency $\eta$ as function of the voltage bias $V$ at fixed level position, $\epsilon = 1.23|U|$, $\Delta T=|U|$, $T=|U|/10$. (b) Efficiency versus power at fixed level position, $\epsilon = 1.23|U|$ for different temperature biases $\Delta T$. (c) Power $\Pmax$ maximized over $V$. (d) Efficiency $\etaPmax$ at maximum power. In all panels, we set $\Gamma_{\text{L}} = \Gamma_{\text{R}} = \Gamma$.
}
\end{figure}

We study the power output $P = I\times V$ with the current $I$ given by \Eq{eq_current} as well as the efficiency $\eta = P/J$, where $J$ is the heat current out of the left (hot) reservoir. Analytical expressions for $J$ are derived in \Ref{Schulenborg2017Dec} for a generic onsite interaction. In \Fig{fig_performance}(a), we show $P$ and $\eta$ as function of $V$ for $\Delta T = |U|$, $\Gamma_{\text{L}} = \Gamma_{\text{R}} = \Gamma$ and for an $\epsilon = 1.23|U|$ in the vicinity of the crossover between single- and two-particle regimes,  optimized towards large power output. The power behaves clearly nonmonotonically with a peak at $|V| \approx \Delta T$. The efficiency $\eta$ \emph{increases} with voltage and assumes about $0.6$ times the Carnot efficiency $\eta_{\text{C}} = 1 - \TR/\TL$ at maximum power. That efficiencies are sizable at finite power output can clearly be seen in \Fig{fig_performance}(b), where $\eta$ and $P$ are shown for several temperature differences $\Delta T$. These also show that efficiencies reach the Carnot limit when power is suppressed at large voltages~\footnote{The sharp feature at low power and high efficiencies stems from the exponential tails of $n_\text{L/R}$ prohibiting an exactly suppressed current. We expect this to be smoothed by here neglected higher-order tunneling effects. }.

Figures~\ref{fig_performance}(c,d) show the power $\Pmax$ maximized over $V$ at otherwise fixed parameters and the efficiency $\etaPmax$ at $\Pmax$. Extended regions of $\epsilon$- and $\Delta T$-values have sizable power $\Pmax \gtrsim \Gamma T$. Interestingly, \Fig{fig_performance}(c) shows that the maximum power is  fully suppressed \textit{only} at $\epsilon = |U|/2$. This can be understood by the fact that the nonlinear thermovoltage $\Snl$ only disappears at this level position, as shown above. However, in the whole two-particle regime, $0<\epsilon<|U|$, the power is small compared to the single-particle regime,  $\epsilon < 0$ and $\epsilon > |U|$. The reason is that the charge relaxation rate $\gamma_\text{R}$ entering the current, Eq.~(\ref{eq_current}), is suppressed in this regime by attractive Coulomb blockade~\cite{Schulenborg2016Feb}. This leads to low power production.

Important for the performance of the attractive quantum dot as a heat engine is our finding that the efficiency at maximum output power is on the order of the Curzon-Ahlborn bound~\cite{Curzon1998Jun}, $\eta_\text{CA}= 1 - \sqrt{\TR/\TL}$, in the whole range in which the output power is sizable. It even reaches this bound close to the \emph{dual} resonances, $\epsilon = 0,|U|$, namely the level positions at which $\nzi$ obtained from the dual mapping changes by 1. It is remarkable that also for the finite power output, prominent features appear at the resonances of the dual repulsive model.

To conclude, we have analyzed the thermoelectric response of a weakly-coupled, single-level quantum dot with attractive interaction together with its performance as a steady-state heat engine. The presented results are expected to be important for future experiments aiming to characterize  systems with strong attractive onsite interaction. At the same time, they demonstrate that nanodevices based on quantum dots with attractive  interaction can also efficiently convert heat to work.

The most relevant qualitative features that we found are:
(i) Two stepped features in the Seebeck coefficient $S$ instead of one, \emph{unexpectedly} located at the positions for resonances of a repulsive dot
[\Fig{fig_linearWB}].
(ii)~Nonlinear Seebeck coefficient $\Snl$ is constant up to sizeable thermal bias [\Fig{fig_nonlinear}].
(iii)~Sharp dip in the maximum power and efficiency at position \emph{expected} for the attractive dot [\Fig{fig_performance}].
We anticipate these effects
in state-of-the-art experiments as, e. g., in quantum dots defined at an oxide interface~\cite{Prawiroatmodjo2017Aug} where clear features of attractive interaction have already been seen in voltage-driven charge transport measurements. Here, applying a temperature bias would allow to verify our predictions, in particular due to the available electrical control over the level position.
In contrast, nanostructures such as in \Ref{Hamo2016Jul} allow the magnitude and sign of the \emph{real} interaction to be altered. This would enable a direct comparison between attractive and \emph{real} repulsive quantum dots.
Predictions for such a comparison are outlined in the Supplementary Information.
However, these setups would need to be extended to allow for transport measurements.

The remarkable appearance of prominent features at level positions characteristic for a repulsive quantum dot was rationalized with a dual mapping emerging from a dissipative symmetry for master equations.
For attractive quantum dots, the role of this dual mapping is particularly important: The dual features with respect to the original attractive system \textit{do not} appear at special positions, where e.g. particle-hole symmetry imposes restrictions. Their prominent role could hence not have been predicted straightforwardly in another way.

\section*{Supplementary material}
See supplementary material for a comparison to a quantum dot with \emph{real} repulsive interaction.

\section*{Acknowledgments and data availability} We acknowledge financial support from the Knut and Alice Wallenberg foundation and the Swedish VR (J.Sp., J.Sc.) and the Danish National Research Foundation (J.Sc.).
The data that supports the findings of this study are available within the article.
\end{document}